\documentclass[prd,aps,floatfix,nofootinbib,11 pt]{revtex4}
\usepackage{amsmath,graphicx,color,epsfig}

\input{tcilatex}

\begin{document}

\title{Quantum Monty Hall problem under decoherence}
\author{Salman Khan\thanks{%
sksafi@phys.qau.edu.pk}, M. Ramzan and M. K. Khan}
\address{Department of Physics Quaid-i-Azam University \\
Islamabad 45320, Pakistan}

\date{\today}

\begin{abstract}
We study the effect of decoherence on quantum Monty Hall problem under the
influence of amplitude damping, depolarizing and dephasing channels. It is
shown that under the effect of decoherence, there is a Nash equilibrium of
the game in case of depolarizing channel for Alice's quantum strategy. Where
as in case of dephasing noise, the game is not influenced by the quantum
channel. For amplitude damping channel, the Bob's payoffs are found
symmetrical with maximum at $p=0.5$ against his classical strategy$.$
However, it is worth-mentioning that in case of depolarizing channel, Bob's
classical strategy remains always dominant against any choice of Alice's
strategy.\newline
\end{abstract}

\maketitle

\
\newline

\vspace*{1.0cm}

\vspace*{1.0cm}


\section{Introduction}

In the recent past, much interest has been developed in the discipline of
quantum information \cite{NieMA} that has led to the creation of quantum
game theory \cite{MeyDA}. Quantum game theory [3-9] has attracted a lot of
attention during the last few years. The quantum Monty Hall problem [9] is
an interesting example in this realm. The quantum game theory has been shown
to be experimentally feasible through the application of a measurement-based
protocol by Prevedel et al. \cite{Prev}. They realized a quantum version of
the Prisoner's Dilemma game based on the entangled photonic cluster states.

In quantum information processing, the main problem is to faithfully
transmit unknown quantum states through a noisy quantum channel. When
quantum information is sent through a channel, the carriers of the
information interact with the channel and get entangled with its many
degrees of freedom. This gives rise to the phenomenon of decoherence on the
state space of the information carriers. Quantum games in the presence of
decoherence have produced interesting results \cite{Dec1,Dec2}. Recently, we
have studied the correlated noise effects in the field of quantum game
theory \cite{Ram}.

In this paper, we study the effect of quantum decoherence on the quantum
Monty Hall problem \cite{MHP}. It is shown that under the effect of
decoherence, a Nash equilibrium of the game exists in case of depolarizing
channel. On the other hand, in case of amplitude damping channel, the Bob's
payoffs are found symmetrical with maximum at $p=0.5$ against his classical
strategy, where $p$ corresponds to the decoherence parameter ranging from $0$
to $1.$ The lower and upper limits of $p$ represents the undecohered and
fully decohered cases respectively. It is also seen that the dephasing noise
have no influence on the game dynamics. However, it is worth-mentioning that
in case of depolarizing channel, Bob's classical strategy becomes dominant
against any choice of Alice's strategy.

\section{Quantum Monty Hall problem}

The well-known classical Monty Hall problem was originally set in the
context of a television game show "\textit{Let's make a deal}" hosted by
Monty Hall. It is a two-person zero sum game involving a prize (car) and
three doors. In the classical version of this problem, the host of the show
(Alice) hides the car behind one of the three closed doors. The guest (Bob)
is asked to select one door out of the three closed doors. Alice then opens
one of the two remaining doors to show that there is no prize behind it.
Then Bob has the option either to stick with his current selection or to
choose the second closed door. This is the actual dilemma of the game.
Classically, switching to the other door increases the winning probability
from one-third to two-third for Bob.

A number of authors have contributed towards the quantization of Monty Hall
problem \cite{MHP,MPH2,Claudia,GM DAriano} They have shown that quantum
entanglement affects the payoffs of the players. In practice no system can
be completely isolated from its environment. Therefore, the interaction
between system and environment leads to the destruction of quantum coherence
of the system. It produces an inevitable noise and results in the loss of
information encoded in the system \cite{Zur}. We proceed with the
quantization protocol of \cite{MHP} and study the effect of decoherence
introduced by the three prototype channels such as amplitude damping,
depolarizing and dephasing channels on the game's dynamics.

We consider that the game space is a 3-dimensional complex Hilbert space
with orthonormal basis $|0\rangle ,$ $|1\rangle $ and $|2\rangle $. Such a
3-dimensional system in Hilbert space is called a qutrit. Alice and Bob
strategies are operators acting on their respective qutrits and are
generally given by $A=a_{ij}$ and $B=b_{ij}$. The open box operator $O,$
which is an open box marking operator and not a measuring operator, is a
unitary operator that can be written as \cite{MHP}

\begin{equation}
O=\sum\limits_{ijkl}\left\vert \varepsilon _{ijk}\right\vert |njk\rangle
\langle ljk|+\sum_{jl}|mjj\rangle \langle ljj|  \label{1}
\end{equation}%
where $\left\vert \varepsilon _{ijk}\right\vert =1,$ if $i,j,k$ are all
different and is $0$ otherwise, $m=\left( j+l+1\right) \ast \left(
mod3\right) $, and $n=\left( i+l\right) \left( mod3\right) $. The second
term of operator $O$ provides an option to Alice for opening an un-chosen
box by Bob. The switching operator $S$ of Bob can be written as

\
\begin{equation}
S=\sum\limits_{ijkl}\left\vert \varepsilon _{ijl}\right\vert |ilk\rangle
\langle ijk|+\sum_{ij}|iij\rangle \langle iij|  \label{2}
\end{equation}%
The second term in equation (\ref{2}) ensures the unitarity of operator $S$
and is irrelevant to the mechanics of the game\cite{MHP}. The Bob's
not-switching operator $N$ is the identity operator $I$ that acts on the
three-qutrit state. The total operator for the not-switching situation of
the game can be written as

\
\begin{equation}
U_{N}=\left( N\sin \gamma \right) O\left( I\otimes B\otimes A\right)
\label{3}
\end{equation}%
and if the Bob switches, the total operator becomes

\
\begin{equation}
U_{S}=\left( S\cos \gamma \right) O\left( I\otimes B\otimes A\right)
\label{E4}
\end{equation}%
where $\gamma \in \left[ 0,\frac{\pi }{2}\right] $, $\gamma =0,$ $\frac{\pi
}{2}$ correspond to the switching and not-switching choices of the Bob
respectively. All the summations in above equations are over the range $0,$ $%
1$ and $2$.

\ The evolution of a state of a quantum system in a noisy environment can be
best described by superoperator $\Phi $ in the Kraus operator representation
as \cite{NieMA}

\
\begin{equation}
\rho _{f}=\Phi \rho _{i}=\sum_{k}E_{k}\rho _{i}E_{k}^{\dag }  \label{E5}
\end{equation}%
where the Kraus operators $E_{i}$ satisfy the following completeness relation

\
\begin{equation}
\sum_{k}E_{k}^{\dag }E_{k}=I  \label{5}
\end{equation}%
The Kraus operators, in our case, for the game are constructed from single
qutrit operators by taking their tensor product over all $n^{2}$ combination
of $\pi \left( i\right) $ indices

\
\begin{equation}
E_{k}=\underset{\pi }{\otimes }e_{\pi \left( i\right) }  \label{6}
\end{equation}%
where $n$ is the number of Kraus operators for a single qutrit channel. The
single qutrit Kraus operators for the amplitude damping channel are given by
\cite{GG}

\
\begin{equation}
E_{0}=\left(
\begin{array}{ccc}
1 & 0 & 0 \\
0 & \sqrt{1-p} & 0 \\
0 & 0 & \sqrt{1-p}%
\end{array}%
\right) ,\ \ E_{1}=\left(
\begin{array}{ccc}
0 & \sqrt{p} & 0 \\
0 & 0 & 0 \\
0 & 0 & 0%
\end{array}%
\right) ,\ \ E_{2}=\left(
\begin{array}{ccc}
0 & 0 & \sqrt{p} \\
0 & 0 & 0 \\
0 & 0 & 0%
\end{array}%
\right)  \label{E7}
\end{equation}%
The Kraus operator for a single qutrit for the dephasing channel are given
by \cite{GG}

\
\begin{equation}
E_{0}=\sqrt{1-p}\left(
\begin{array}{ccc}
1 & 0 & 0 \\
0 & 1 & 0 \\
0 & 0 & 1%
\end{array}%
\right) ,\ \ E_{1}=\sqrt{p}\left(
\begin{array}{ccc}
1 & 0 & 0 \\
0 & \omega & 0 \\
0 & 0 & \omega ^{2}%
\end{array}%
\right) ,  \label{7}
\end{equation}%
where $p$ corresponds to the decoherence parameter and $\omega =e^{2i\pi /3}$%
. The single qutrit Kraus operators for the depolarizing channel are given
by \cite{Berg}

\
\begin{equation}
E_{0}=\sqrt{1-p}I,\ E_{1}=\sqrt{\frac{p}{8}}Y,\ E_{2}=\sqrt{\frac{p}{8}}Z,\
E_{3}=\sqrt{\frac{p}{8}}Y^{2},\ E_{4}=\sqrt{\frac{p}{8}}YZ
\end{equation}

\
\begin{equation}
E_{5}=\sqrt{\frac{p}{8}}Y^{2}Z,\ E_{6}=\sqrt{\frac{p}{8}}YZ^{2},\ \ E_{7}=%
\sqrt{\frac{p}{8}}Y^{2}Z^{2},\ \ E_{8}=\sqrt{\frac{p}{8}}Z^{2}  \label{E8}
\end{equation}%
where

\
\begin{equation}
Y=\left(
\begin{array}{ccc}
0 & 1 & 0 \\
0 & 0 & 1 \\
1 & 0 & 0%
\end{array}%
\right) ,\ \ Z=\left(
\begin{array}{ccc}
1 & 0 & 0 \\
0 & \omega & 0 \\
0 & 0 & \omega ^{2}%
\end{array}%
\right)  \label{9}
\end{equation}%
We consider the following maximally entangled qutrit state, which is shared
between Alice and Bob

\
\begin{equation}
|\psi _{i}\rangle =|0\rangle \otimes \frac{1}{\sqrt{3}}\left( |00\rangle
+|11\rangle +|22\rangle \right)  \label{10}
\end{equation}%
In equation (\ref{10}) the state $|0\rangle $ stands for the open door. The
final state for the case when Bob does not switch to the other door becomes

\
\begin{equation}
\rho _{fN}=U_{N}\left( \sum_{k}E_{k}\rho _{i}E_{k}^{\dag }\right)
U_{N}^{\dag }  \label{11}
\end{equation}%
For the case of switching to the other door, the game final state becomes

\
\begin{equation}
\rho _{fS}=U_{S}\left( \sum_{k}E_{k}\rho _{i}E_{k}^{\dag }\right)
U_{S}^{\dag }  \label{12}
\end{equation}%
where
\begin{equation}
\rho _{i}=|\psi _{i}\rangle \left\langle \psi _{i}\right\vert  \label{13}
\end{equation}%
Bob wins if he chooses the door behind which the prize is located. Hence the
payoff of Bob is given by

\
\begin{equation}
\left\langle \$_{B}\right\rangle =\left\langle \$_{B}\right\rangle
_{N}+\left\langle \$_{B}\right\rangle _{S}=\sum_{ijj}\left( \left( \rho
_{fN}\right) _{ijj}+\left( \rho _{fS}\right) _{ijj}\right)  \label{14}
\end{equation}%
The payoff of Alice is then given by $\left\langle \$_{A}\right\rangle
=1-\left\langle \$_{B}\right\rangle .$

\section{Calculation, results and discussions}

In this section, we present the results of our calculations based on the
three prototype channels such as amplitude damping, depolarizing and
dephasing channels parametrized by the decoherence parameter $p$.

\subsection{Amplitude damping channel}

By using equations (\ref{E7}, \ref{10} \ and \ref{13}), the Bob's payoff for
not-switching case can be written as

\
\begin{eqnarray}
\left\langle \$_{B}\right\rangle _{N} &=&-\frac{1}{3}%
(A_{1}(-1-2p^{2})+A_{2}(-1+2p-p^{2})  \notag \\
&&+(-1+p)(A_{3}p+(a_{01}b_{01}+a_{02}b_{02})a_{00}^{\ast }b_{00}^{\ast
}+(a_{00}b_{00}+a_{02}b_{02}  \notag \\
&&-a_{02}b_{02}p)a_{01}^{\ast }b_{01}^{\ast
}+(a_{00}b_{00}+a_{01}b_{01}-a_{01}b_{01}p)a_{02}^{\ast }b_{02}^{\ast }
\notag \\
&&+(a_{11}b_{11}+a_{12}b_{12})a_{10}^{\ast }b_{10}^{\ast
}+(a_{10}b_{10}+a_{12}b_{12}-a_{12}b_{12}p)a_{11}^{\ast }b_{11}^{\ast }
\notag \\
&&+(a_{10}b_{10}+a_{11}b_{11}-a_{11}b_{11}p)a_{12}^{\ast }b_{12}^{\ast }
\notag \\
&&+(a_{21}b_{21}+a_{22}b_{22})a_{20}^{\ast }b_{20}^{\ast
}+(a_{20}b_{20}+a_{22}b_{22}-a_{22}b_{22}p)a_{21}^{\ast }b_{21}^{\ast }
\notag \\
&&+(a_{20}b_{20}+a_{21}b_{21}-a_{21}b_{21}p)a_{22}^{\ast }b_{22}^{\ast
})\sin ^{2}\gamma  \label{15}
\end{eqnarray}

\ where the coefficients $A_{i}$ are given in appendix A.

\ For the case when Bob switches to the other door, the payoff can be
obtained by using equations (\ref{E7}, \ref{12} and \ref{13}) which reads

\
\begin{eqnarray}
\left\langle \$_{B}\right\rangle _{S} &=&-\frac{1}{3}%
((-1-2p^{2})B_{1}+(-1+p)pB_{2}+(-1+p)^{2}B_{3}  \notag \\
&&+(-a_{12}b_{02}a_{11}^{\ast }-a_{22}b_{02}a_{21}^{\ast })b_{01}^{\ast
}-(a_{11}b_{01}a_{12}^{\ast }b+a_{21}b_{01}a_{22}^{\ast })b_{02}^{\ast }
\notag \\
&&-(a_{02}b_{12}a_{01}^{\ast }+a_{22}b_{12}a_{21}^{\ast })b_{11}^{\ast
}-(a_{01}b_{11}a_{02}^{\ast }+a_{21}b_{11}a_{22}^{\ast })b_{12}^{\ast }
\notag \\
&&-(a_{02}b_{22}a_{01}^{\ast }+a_{12}b_{22}a_{11}^{\ast })b_{21}^{\ast
}-(a_{01}b_{21}a_{02}^{\ast }+a_{11}b_{21}a_{12}^{\ast })b_{22}^{\ast })
\notag \\
&&+(-1+p)(a_{20}^{\ast }((a_{21}b_{01}+a_{22}b_{02})b_{00}^{\ast
}+(a_{21}b_{11}+a_{22}b_{12})b_{10}^{\ast })  \notag \\
&&+a_{20}a_{21}^{\ast }(b_{00}b_{01}^{\ast }+b_{10}b_{11}^{\ast
})+a_{20}a_{22}^{\ast }(b_{00}b_{02}^{\ast }+b_{10}b_{12}^{\ast })  \notag \\
&&+a_{00}^{\ast }((a_{01}b_{11}+a_{02}b_{12})b_{10}^{\ast
}+(a_{01}b_{21}+a_{02}b_{22})b_{20}^{\ast })  \notag \\
&&+a_{10}^{\ast }((a_{11}b_{01}+a_{12}b_{02})b_{00}^{\ast
}+(a_{11}b_{21}+a_{12}b_{22})b_{20}^{\ast })  \notag \\
&&+a_{10}a_{11}^{\ast }(b_{00}b_{01}^{\ast }+b_{20}b_{21}^{\ast
})+a_{10}a_{12}^{\ast }(b_{00}b_{02}^{\ast }+b_{20}b_{22}^{\ast })  \notag \\
&&+a_{00}(a_{01}^{\ast }(b_{10}b_{11}^{\ast }+b_{20}b_{21}^{\ast
})+a_{02}^{\ast }(b_{10}b_{12}^{\ast }+b_{20}b_{22}^{\ast }))))\cos
^{2}\gamma  \label{16}
\end{eqnarray}%
Where $B_{i}$,are given in appendix A. The total payoff of Bob is the sum of
equations (\ref{15} and \ref{16}).

\ To analyze our results, we consider that let Bob has access to a classical
strategy only (i.e., $B=I$), therefore, he can select any door with equal
probability. Then Bob's total payoff becomes

\
\begin{eqnarray}
\left\langle \$_{B}\right\rangle &=&-\frac{1}{3}(-(\left\vert
a_{01}\right\vert ^{2}+\left\vert a_{02}\right\vert ^{2}+\left\vert
a_{12}\right\vert ^{2}+\left\vert a_{21}\right\vert ^{2})(-1+p)^{2}  \notag
\\
&&+(2\left\vert a_{00}\right\vert ^{2}+\left\vert a_{10}\right\vert
^{2}+\left\vert a_{11}\right\vert ^{2}+\left\vert a_{12}\right\vert
^{2}+\left\vert a_{20}\right\vert ^{2}+\left\vert a_{21}\right\vert ^{2}
\notag \\
&&+\left\vert a_{22}\right\vert ^{2})(-1+p)p+(\left\vert a_{10}\right\vert
^{2}+\left\vert a_{20}\right\vert ^{2})(-1-2p^{2}))\cos ^{2}\gamma  \notag \\
&&-1/3(\left\vert a_{00}\right\vert ^{2}(-1-2p^{2})+(\left\vert
a_{11}\right\vert ^{2}+\left\vert a_{22}\right\vert ^{2})(-1+2p-p^{2})
\notag \\
&&+(\left\vert a_{01}\right\vert ^{2}+\left\vert a_{02}\right\vert
^{2}+\left\vert a_{10}\right\vert ^{2}+\left\vert a_{20}\right\vert
^{2})(-p+p^{2}))\sin ^{2}\gamma  \label{17}
\end{eqnarray}%
Now, Alice can make the game fair if she uses an operator whose every
diagonal element has an absolute value of $\frac{1}{\sqrt{2}}$ and every
off-diagonal element has an absolute value of $\frac{1}{2}$. One such SU$%
\left( 3\right) $ operator is

\
\begin{equation}
H=\left(
\begin{array}{ccc}
\frac{1}{\sqrt{2}} & \frac{1}{2} & \frac{1}{2} \\
-\frac{1}{2} & \frac{3-i\sqrt{7}}{4\sqrt{2}} & \frac{1+i\sqrt{7}}{4\sqrt{2}}
\\
\frac{-1-i\sqrt{7}}{4\sqrt{2}} & \frac{-3+i\sqrt{7}}{8} & \frac{5+i\sqrt{7}}{%
8}%
\end{array}%
\right)  \label{18}
\end{equation}%
The total payoff for Bob by using the above operator is obtained as

\
\begin{equation}
\left\langle \$_{B}\right\rangle =\frac{1}{6}\left( (3-2(-1+p)p)\cos
^{2}\gamma +(3+2(-1+p)p)\sin ^{2}\gamma \right)  \label{19}
\end{equation}

It is easy to check that by setting $p=0,$ in equation (\ref{19}), the Bob's
payoff reduces to the results obtained in ref. \cite{MHP}. For $p=0.5$, the
probability of Bob to win increases to $58.33\%$ if he changes the current
selection and switch to the other door. This result for the Bob's payoff is
different both from the classical result $(66\%)$ and quantum mechanical
result $(50\%)$ without decoherence. If Bob sticks to his current selection,
his winning probability is $41.66\%$, lesser than the quantum mechanical
payoff $\frac{1}{2}$. The dependence of Bob's payoff on $p$ for both
switching and not-switching cases is shown in figure 1. However, instead of
classical strategy (identity operator) if Bob has access to the quantum
strategy too, that is, if Bob uses any one of the following operators

\
\begin{equation}
M_{1}=\left(
\begin{array}{ccc}
0 & 1 & 0 \\
0 & 0 & 1 \\
1 & 0 & 0%
\end{array}%
\right) \ \ \ \ \ \ \ \ \ \ \ \ \ \ \ \ \ \ M_{2}=\left(
\begin{array}{ccc}
0 & 0 & 1 \\
1 & 0 & 0 \\
0 & 1 & 0%
\end{array}%
\right)  \label{E19}
\end{equation}%
and then switches, his payoff increases. The dependence of Bob's payoff for $%
B=M_{i}$ is shown in figure 3.

Now consider the situation where Alice is restricted to a classical
strategy, that is, Alice operates $A=I$. Then, in the undecohered case, Bob
wins $(\$_{B}=1)$ by using $B=I,$ if he does not switch. However, under
decoherence, Bob's winning probability reduces to two-third for $p=0.5$ even
if he does not switch. If Bob switches, his winning probability reduces to
one-third for $p=0.5.$ Similarly, for $p=0$ Bob wins if he uses $M_{i}$ $%
(M_{1}$ or $M_{2})$ and then switches. Where as in the presence of noise,
Bob's winning probability reduces to $83.3\%$ for $p=0.5$ even if he
switches. The dependence of Bob's payoff on decoherence parameter $p$ for
the case when both Alice and Bob use classical strategies is shown in figure
2. For Alice to make the game fair $($i.e. $A=H)$, the maximum value of
Bob's payoff occurs if he uses either of $M_{i}$ and then switches. In
conclusion, the Bob's payoffs are found symmetrical with maximum at $p=0.5$
against his classical strategy (see figure 3)$.$

\subsection{Depolarizing channel}

In case of depolarizing channel, the Bob's payoff for not-switching case can
be found by using equations (\ref{11} and \ref{E8}) as

\
\begin{eqnarray}
\left\langle \$_{B}\right\rangle _{N} &=&-\frac{1}{192}%
(C-(8-9p)^{2}((a_{01}b_{01}+a_{02}b_{02})a_{00}^{\ast }b_{00}^{\ast
}+(a_{00}b_{00}+a_{02}b_{02})a_{01}^{\ast }b_{01}^{\ast }  \notag \\
&&+(a_{00}b_{00}+a_{01}b_{01})a_{02}^{\ast }b_{02}^{\ast
}+(a_{11}b_{11}+a_{12}b_{12})a_{10}^{\ast }b_{10}^{\ast
}+(a_{10}b_{10}+a_{12}b_{12})a_{11}^{\ast }b_{11}^{\ast }  \notag \\
&&+(a_{10}b_{10}+a_{11}b_{11})a_{12}^{\ast }b_{12}^{\ast
}+(a_{21}b_{21}+a_{22}b_{22})a_{20}^{\ast }b_{20}^{\ast }  \notag \\
&&+(a_{20}b_{20}+a_{22}b_{22})a_{21}^{\ast }b_{21}^{\ast
}+(a_{20}b_{20}+a_{21}b_{21})a_{22}^{\ast }b_{22}^{\ast }))\sin ^{2}\gamma
\label{20}
\end{eqnarray}%
where $C$ is given in appendix A. Similarly, using equations (\ref{E8}, \ref%
{12} and \ref{13}), the Bob's payoff for the switching case becomes

\
\begin{eqnarray}
\left\langle \$_{B}\right\rangle _{S} &=&\frac{1}{192}%
((48p-27p^{2})D_{1}+(64-96p+54p^{2})D_{2}  \notag \\
&&+(8-9p)^{2}((a_{21}b_{01}+a_{22}b_{02})a_{20}^{\ast }b_{00}^{\ast
}+(a_{20}b_{00}+a_{22}b_{02})a_{21}^{\ast }b_{01}^{\ast }  \notag \\
&&+(a_{20}b_{00}+a_{21}b_{01})a_{22}^{\ast }b_{02}^{\ast
}+(a_{21}b_{11}+a_{22}b_{12})a_{20}^{\ast }b_{10}^{\ast }  \notag \\
&&+(a_{20}b_{10}+a_{22}b_{12})a_{21}^{\ast }b_{11}^{\ast
}+(a_{20}b_{10}+a_{21}b_{11})a_{22}^{\ast }b_{12}^{\ast }  \notag \\
&&+(a_{01}b_{11}+a_{02}b_{12})a_{00}^{\ast }b_{10}^{\ast
}+(a_{01}b_{21}+a_{02}b_{22})a_{00}^{\ast }b_{20}^{\ast }  \notag \\
&&+(a_{11}b_{01}+a_{12}b_{02})a_{10}^{\ast }b0_{0}^{\ast
}+(a_{11}b_{21}+a_{12}b_{22})a_{10}^{\ast }b_{20}^{\ast }  \notag \\
&&+(a_{00}b_{10}+a_{02}b_{12})a_{01}^{\ast }b_{11}^{\ast
}+(a_{00}b_{20}+a_{02}b_{22})a_{01}^{\ast }b_{21}^{\ast }  \notag \\
&&+(a_{10}b_{00}+a_{12}b_{02})a_{11}^{\ast }b_{01}^{\ast
}+(a_{10}b_{20}+a_{12}b_{22})a_{11}^{\ast }b_{21}^{\ast }  \notag \\
&&+(a_{00}b_{10}+a0_{1}b_{11})a_{02}^{\ast }b_{12}^{\ast
}+(a_{00}b_{20}+a_{01}b_{21})a_{02}^{\ast }b_{22}^{\ast }  \notag \\
&&+(a_{10}b_{00}+a_{11}b_{01})a_{12}^{\ast }b_{02}^{\ast
}+(a_{10}b_{20}+a_{11}b_{21})a_{12}^{\ast }b_{22}^{\ast }))\cos ^{2}\gamma
\label{21}
\end{eqnarray}%
where $D_{i}$ are given in appendix A.

\ Bob's total payoff is the sum of equations (\ref{20} and \ref{21}). To
analyze the effect of decoherence on Bob's payoff we consider for example
that Bob plays the classical strategy $B=I,$ then Bob's total payoff becomes

\
\begin{eqnarray}
\left\langle \$_{B}\right\rangle &=&1/192((64-96p+54p^{2})(\left\vert
a_{01}\right\vert ^{2}+\left\vert a_{02}\right\vert ^{2}+\left\vert
a_{10}\right\vert ^{2}+\left\vert a_{12}\right\vert ^{2}  \notag \\
&&+\left\vert a_{20}\right\vert ^{2}+\left\vert a_{21}\right\vert
^{2})+(48p-27p^{2})(2\left\vert a_{00}\right\vert ^{2}+\left\vert
a_{01}\right\vert ^{2}+\left\vert a_{02}\right\vert ^{2}  \notag \\
&&+\left\vert a_{10}\right\vert ^{2}+2\left\vert a_{11}\right\vert
^{2}+\left\vert a_{12}\right\vert ^{2}+\left\vert a_{20}\right\vert
^{2}+\left\vert a_{21}\right\vert ^{2}+2\left\vert a_{22}\right\vert
^{2})\cos ^{2}\gamma  \notag \\
&&-((-48p+27p^{2})(\left\vert a_{01}\right\vert ^{2}+\left\vert
a_{02}\right\vert ^{2}+\left\vert a_{10}\right\vert ^{2}+\left\vert
a_{12}\right\vert ^{2}+\left\vert a_{20}\right\vert ^{2}+\left\vert
a_{21}\right\vert ^{2})  \notag \\
&&+(-64+96p-54p^{2})(\left\vert a_{00}\right\vert ^{2}+\left\vert
a_{11}\right\vert ^{2}+\left\vert a_{22}\right\vert ^{2}))\sin ^{2}\gamma )
\label{22}
\end{eqnarray}%
The game becomes fair if Alice uses the operator $H$ as given in equation (%
\ref{18}). Bob's total payoff then becomes

\
\begin{equation}
\left\langle \$_{B}\right\rangle =\frac{1}{128}((64+3(16-9p)p)\cos
^{2}\gamma +(64+3p(-16+9p))\sin ^{2}\gamma )  \label{23}
\end{equation}

We can see that in the absence of decoherence i.e. $p=0$, our results
equation (\ref{23}) reduces to the results of ref. \cite{MHP}. However, in
the presence of decoherence i.e. for $p=0.5$, Bob's winning probability
increases to $63.47\%$ instead of $50\%$ if he switches to the other door
and if he sticks to his current selection, the winning probability is just $%
36.52\%.$ On the other hand, for $p=0.9$ the classical results are
retrieved. The Bob's payoffs for $A=H\,$and $B=I$ in the presence of
decoherence for both switching and not-switching cases are shown in figure 4.

Bob's payoffs in the presence of decoherence for $B=A=I,$ are shown in
figure 5. We see from figure 5 that Bob can win with a two-third probability
for $p=0.9$ even if he switches to the other door. This is in contrary to
the $p=0$ situation, where Bob loses $(\left\langle \$_{B}\right\rangle =0)$
in the case of switching to the other door. Thus, it gives rise to the Nash
equilibrium of the game under decoherence. Further more, If Bob uses $M_{i}$
and then switches, his winning probability varies from $1$ to $\frac{2}{3}$
as $p$ varies from $0$ to $1$.

\subsection{Dephasing Channel}

Bob's payoff for not-switching case in the presence of dephasing noise can
be written, by using equations (\ref{11}, \ref{13}) and \ref{7}), as

\
\begin{eqnarray}
\left\langle \$_{B}\right\rangle _{N} &=&\frac{1}{6}%
(2E+2((a_{01}b_{01}+a_{02}b_{02})a_{00}^{\ast }b_{00}^{\ast
}+(a_{00}b_{00}+a_{02}b_{02})a_{01}^{\ast }b_{01}^{\ast }  \notag \\
&&+(a_{00}b_{00}+a_{01}b_{01})a_{02}^{\ast }b_{02}^{\ast
}+(a_{11}b_{11}+a_{12}b_{12})a_{10}^{\ast }b_{10}^{\ast }  \notag \\
&&+(a_{10}b_{10}+a_{12}b_{12})a_{11}^{\ast }b_{11}^{\ast
}+(a_{10}b_{10}+a_{11}b_{11})a_{12}^{\ast }b_{12}^{\ast }  \notag \\
&&+(a_{21}b_{21}+a_{22}b_{22})a_{20}^{\ast }b_{20}^{\ast
}+(a_{20}b_{20}+a_{22}b_{22})a_{21}^{\ast }b_{21}^{\ast }  \notag \\
&&+(a_{20}b_{20}+a_{21}b_{21})a_{22}^{\ast }b_{22}^{\ast
})+3(-2+p)p((a_{01}b_{01}+a_{02}b_{02})a_{00}^{\ast }b_{00}^{\ast }  \notag
\\
&&+(a_{00}b_{00}+a_{02}b_{02})a_{01}^{\ast }b_{01}^{\ast
}+(a_{00}b_{00}+a_{01}b_{01})a_{02}^{\ast }b_{02}^{\ast }  \notag \\
&&+(a_{11}b_{11}+a_{12}b_{12})a_{10}^{\ast }b_{10}^{\ast
}+(a_{10}b_{10}+a_{12}b_{12})a_{11}^{\ast }b_{11}^{\ast }  \notag \\
&&+(a_{10}b_{10}+a_{11}b_{11})a_{12}^{\ast }b_{12}^{\ast
}+(a_{21}b_{21}+a_{22}b_{22})a_{20}^{\ast }b_{20}^{\ast }  \notag \\
&&+(a_{20}b_{20}+a_{22}b_{22})a_{21}b_{21}+(a_{20}b_{20}+a_{21}b_{21})a_{22}^{\ast }b_{22}^{\ast })
\notag \\
&&+\sqrt{3}ip(-2+3p)(-(a_{01}b_{01}-a_{02}b_{02})a_{00}^{\ast }b_{00}^{\ast
}+(a_{00}b_{00}-a_{02}b_{02})a_{01}^{\ast }b_{01}^{\ast }  \notag \\
&&-(a_{00}b_{00}-a_{01}b_{01})a_{02}^{\ast }b_{02}^{\ast
}-(a_{11}b_{11}-a_{12}b_{12})a_{10}^{\ast }b_{10}^{\ast }  \notag \\
&&+(a_{10}b_{10}-a_{12}b_{12})a_{11}^{\ast }b_{11}^{\ast
}-(a_{10}b_{10}-a_{11}b_{11})a_{12}^{\ast }b_{12}^{\ast
}-(a_{21}b_{21}-a_{22}b_{22})a_{20}^{\ast }b_{20}^{\ast }  \notag \\
&&+(a_{20}b_{20}-a_{22}b_{22})a_{21}^{\ast }b_{21}^{\ast
}-(a_{20}b_{20}-a_{21}b21)a_{22}^{\ast }b_{22}^{\ast }))\sin ^{2}\gamma
\label{24}
\end{eqnarray}%
where $E$ is given in appendix A. However, if Bob switches to the other door
his payoff becomes

\
\begin{eqnarray}
\left\langle \$_{B}\right\rangle _{S} &=&\frac{1}{6}(2F_{1}+3(1+\sqrt{3}%
i)p^{2}F_{2}+(3-3\sqrt{3}i)p^{2}F_{3}  \notag \\
&&+(2+2(-3+\sqrt{3}i)p)F_{4}+(2-2(3+\sqrt{3}i)p)F_{5})\cos ^{2}\gamma
\label{25}
\end{eqnarray}%
Where $F_{i}$ are given in appendix A. It is important to note here that
Bob's payoffs for not-switching and switching cases are written in general
form. Bob's total payoff reduces to the result of ref. \cite{MHP} when we
set the decoherence parameter $p=0$ in the general relation. Further more,
when either Bob or Alice is restricted to a classical strategy the total
Bob's payoff becomes independent of decoherence parameter $p$. The same is
true for the case of when Alice and Bob use quantum strategies. Therefore,
the dephasing noise does not influence the game.

\section{Conclusions}

We study the quantum Monty Hall problem under the influence of amplitude
damping, depolarizing and dephasing channels. A Nash equilibrium of the game
exists under the effect of decoherence against Alice's quantum strategy in
the case of depolarizing channel. It is also seen that the dephasing noise
does not influence the game in contrary to the depolarizing and amplitude
damping channels. It is worth-mentioning that for amplitude damping and
depolarizing channels, Bob's classical strategy is superior over any choice
of Alice's strategy.

{\Large Appendix A}\newline
The coefficients $A_{i}$ in equation (\ref{15}) are given below,\
\begin{eqnarray}
A_{1} &=&\left\vert a_{00}\right\vert ^{2}\left\vert b_{00}\right\vert
^{2}+\left\vert a_{10}\right\vert ^{2}\left\vert b_{10}\right\vert
^{2}+\left\vert a_{20}\right\vert ^{2}\left\vert b_{20}\right\vert ^{2}
\notag \\
A_{2} &=&\left\vert a_{01}\right\vert ^{2}\left\vert b_{01}\right\vert
^{2}+\left\vert a_{02}\right\vert ^{2}\left\vert b_{02}\right\vert
^{2}+\left\vert a_{11}\right\vert ^{2}\left\vert b_{11}\right\vert
^{2}+\left\vert a_{12}\right\vert ^{2}\left\vert b_{12}\right\vert ^{2}
\notag \\
&&+\left\vert a_{21}\right\vert ^{2}\left\vert b_{21}\right\vert
^{2}+\left\vert a_{22}\right\vert ^{2}\left\vert b_{22}\right\vert ^{2}
\notag \\
A_{3} &=&\left\vert a_{01}\right\vert ^{2}\left\vert b_{00}\right\vert
^{2}+\left\vert a_{02}\right\vert ^{2}\left\vert b_{00}\right\vert
^{2}+\left\vert a_{00}\right\vert ^{2}\left\vert b_{01}\right\vert
^{2}+\left\vert a_{00}\right\vert ^{2}\left\vert b_{02}\right\vert ^{2}
\notag \\
&&+\left\vert a_{11}\right\vert ^{2}\left\vert b_{10}\right\vert
^{2}+\left\vert a_{12}\right\vert ^{2}\left\vert b_{10}\right\vert
^{2}+\left\vert a_{10}\right\vert ^{2}\left\vert b_{11}\right\vert
^{2}+\left\vert a_{10}\right\vert ^{2}\left\vert b_{12}\right\vert ^{2}
\notag \\
&&+\left\vert a_{21}\right\vert ^{2}\left\vert b_{20}\right\vert
^{2}+\left\vert a_{22}\right\vert ^{2}\left\vert b_{20}\right\vert
^{2}+\left\vert a_{20}\right\vert ^{2}\left\vert b_{21}\right\vert
^{2}+\left\vert a_{20}\right\vert ^{2}\left\vert b_{22}\right\vert ^{2}
\end{eqnarray}%
The coefficients$\ B_{i}$ in equation (\ref{16}) are given below,\
\begin{eqnarray}
B_{1} &=&\left\vert a_{20}\right\vert ^{2}(\left\vert b_{00}\right\vert
^{2}+\left\vert b_{10}\right\vert ^{2})+\left\vert a_{10}\right\vert
^{2}(\left\vert b_{00}\right\vert ^{2}+\left\vert b_{20}\right\vert
^{2})+\left\vert a_{00}\right\vert ^{2}(\left\vert b_{10}\right\vert
^{2}+\left\vert b_{20}\right\vert ^{2})  \notag \\
B_{2} &=&\left\vert a_{21}\right\vert ^{2}\left\vert b_{00}\right\vert
^{2}+\left\vert a_{22}\right\vert ^{2}\left\vert b_{00}\right\vert
^{2}+\left\vert a_{10}\right\vert ^{2}\left\vert b_{01}\right\vert
^{2}+\left\vert a_{20}\right\vert ^{2}\left\vert b_{01}\right\vert
^{2}+\left\vert a_{10}\right\vert ^{2}\left\vert b_{02}\right\vert
^{2}+\left\vert a_{20}\right\vert ^{2}\left\vert b_{02}\right\vert ^{2}
\notag \\
&&+\ \left\vert a_{01}\right\vert ^{2}\left\vert b_{10}\right\vert
^{2}+\left\vert a_{02}\right\vert ^{2}\left\vert b_{10}\right\vert
^{2}+\left\vert a_{21}\right\vert ^{2}\left\vert b_{10}\right\vert
^{2}+\left\vert a_{22}\right\vert ^{2}\left\vert b_{10}\right\vert ^{2}
\notag \\
&&+\left\vert a_{00}\right\vert ^{2}\left\vert b_{11}\right\vert
^{2}+\left\vert a_{20}\right\vert ^{2}\left\vert b_{11}\right\vert
^{2}+\left\vert a_{00}\right\vert ^{2}\left\vert b_{12}\right\vert
^{2}+\left\vert a_{20}\right\vert ^{2}\left\vert b_{12}\right\vert ^{2}
\notag \\
&&+\left\vert a_{01}\right\vert ^{2}\left\vert b_{20}\right\vert
^{2}+\left\vert a_{02}\right\vert ^{2}\left\vert b_{20}\right\vert
^{2}+\left\vert a_{11}\right\vert ^{2}(\left\vert b_{00}\right\vert
^{2}+\left\vert b_{20}\right\vert ^{2})  \notag \\
&&+\left\vert a_{12}\right\vert ^{2}(\left\vert b_{00}\right\vert
^{2}+\left\vert b_{20}\right\vert ^{2})+\left\vert a_{00}\right\vert
^{2}\left\vert b_{21}\right\vert ^{2}+\left\vert a_{10}\right\vert
^{2}\left\vert b_{21}\right\vert ^{2}+\left\vert a_{00}\right\vert
^{2}\left\vert b_{22}\right\vert ^{2}  \notag \\
&&+\left\vert a_{10}\right\vert ^{2}\left\vert b_{22}\right\vert ^{2}  \notag
\\
B_{3} &=&\left\vert a_{12}\right\vert ^{2}\left\vert b_{02}\right\vert
^{2}+\left\vert a_{22}\right\vert ^{2}\left\vert b_{02}\right\vert
^{2}+\left\vert a_{01}\right\vert ^{2}\left\vert b_{11}\right\vert ^{2}
\notag \\
&&+\left\vert a_{21}\right\vert ^{2}(\left\vert b_{01}\right\vert
^{2}+\left\vert b_{11}\right\vert ^{2})+\left\vert a_{02}\right\vert
^{2}\left\vert b_{12}\right\vert ^{2}+\left\vert a_{22}\right\vert
^{2}\left\vert b_{12}\right\vert ^{2}+\left\vert a_{01}\right\vert
^{2}\left\vert b_{21}\right\vert ^{2}  \notag \\
&&+\left\vert a_{11}\right\vert ^{2}(\left\vert b_{01}\right\vert
^{2}+\left\vert b_{21}\right\vert ^{2})+\left\vert a_{02}\right\vert
^{2}\left\vert b_{22}\right\vert ^{2}+\left\vert a_{12}\right\vert
^{2}\left\vert b_{22}\right\vert ^{2}
\end{eqnarray}%
The coefficient$\ C$ in equation (\ref{20}) is given below,\
\begin{eqnarray}
C &=&\left\vert a_{00}\right\vert ^{2}\left\vert b_{00}\right\vert
^{2}+\left\vert a_{01}\right\vert ^{2}\left\vert b_{01}\right\vert
^{2}+\left\vert a_{02}\right\vert ^{2}\left\vert b_{02}\right\vert
^{2}+\left\vert a_{10}\right\vert ^{2}\left\vert b_{10}\right\vert ^{2}
\notag \\
&&+\left\vert a_{11}\right\vert ^{2}\left\vert b_{11}\right\vert
^{2}+\left\vert a_{12}\right\vert ^{2}\left\vert b_{12}\right\vert
^{2}+\left\vert a_{20}\right\vert ^{2}\left\vert b_{20}\right\vert
^{2}+\left\vert a_{21}\right\vert ^{2}\left\vert b_{21}\right\vert ^{2}
\notag \\
&&+\left\vert a_{22}\right\vert ^{2}\left\vert b_{22}\right\vert
^{2})+(-48p+27p^{2})(\left\vert a_{01}\right\vert ^{2}\left\vert
b_{00}\right\vert ^{2}+\left\vert a_{02}\right\vert ^{2}\left\vert
b_{00}\right\vert ^{2}  \notag \\
&&+\left\vert a_{00}\right\vert ^{2}\left\vert b_{01}\right\vert
^{2}+\left\vert a_{02}\right\vert ^{2}\left\vert b_{01}\right\vert
^{2}+\left\vert a_{00}\right\vert ^{2}\left\vert b_{02}\right\vert
^{2}+\left\vert a_{01}\right\vert ^{2}\left\vert b_{02}\right\vert ^{2}
\notag \\
&&+\left\vert a_{11}\right\vert ^{2}\left\vert b_{10}\right\vert
^{2}+\left\vert a_{12}\right\vert ^{2}\left\vert b_{10}\right\vert
^{2}+\left\vert a_{10}\right\vert ^{2}\left\vert b_{11}\right\vert
^{2}+\left\vert a_{12}\right\vert ^{2}\left\vert b_{11}\right\vert ^{2}
\notag \\
&&+\left\vert a_{10}\right\vert ^{2}\left\vert b_{12}\right\vert
^{2}+\left\vert a_{11}\right\vert ^{2}\left\vert b_{12}\right\vert
^{2}+\left\vert a_{21}\right\vert ^{2}\left\vert b_{20}\right\vert
^{2}+\left\vert a_{22}\right\vert ^{2}\left\vert b_{20}\right\vert ^{2}
\notag \\
&&+\left\vert a_{20}\right\vert ^{2}\left\vert b_{21}\right\vert
^{2}+\left\vert a_{22}\right\vert ^{2}\left\vert b_{21}\right\vert
^{2}+\left\vert a_{20}\right\vert ^{2}\left\vert b_{22}\right\vert
^{2}+\left\vert a_{21}\right\vert ^{2}\left\vert b_{22}\right\vert ^{2}
\end{eqnarray}%
The coefficients$\ D_{i}$ in equation (\ref{21}) are given below,

\begin{eqnarray}
D_{1} &=&\left\vert a_{21}\right\vert ^{2}\left\vert b_{00}\right\vert
^{2}+\left\vert a_{22}\right\vert ^{2}\left\vert b_{00}\right\vert
^{2}+\left\vert a_{10}\right\vert ^{2}|\left\vert b_{01}\right\vert
^{2}+\left\vert a_{20}\right\vert ^{2}\left\vert b_{01}\right\vert ^{2}
\notag \\
&&+\left\vert a_{22}\right\vert ^{2}\left\vert b_{01}\right\vert
^{2}+\left\vert a_{10}\right\vert ^{2}\left\vert b_{02}\right\vert
^{2}+\left\vert a_{20}\right\vert ^{2}\left\vert b_{02}\right\vert
^{2}+\left\vert a_{21}\right\vert ^{2}\left\vert b_{02}\right\vert ^{2}
\notag \\
&&+\left\vert a_{01}\right\vert ^{2}\left\vert b_{10}\right\vert
^{2}+\left\vert a_{02}\right\vert ^{2}\left\vert b_{10}\right\vert
^{2}+\left\vert a_{21}\right\vert ^{2}\left\vert b_{10}\right\vert
^{2}+\left\vert a_{22}\right\vert ^{2}\left\vert b_{10}\right\vert ^{2}
\notag \\
&&+\left\vert a_{00}\right\vert ^{2}\left\vert b_{11}\right\vert
^{2}+\left\vert a_{02}\right\vert ^{2}\left\vert b_{11}\right\vert
^{2}+\left\vert a_{20}\right\vert ^{2}\left\vert b_{11}\right\vert
^{2}+\left\vert a_{22}\right\vert ^{2}\left\vert b_{11}\right\vert ^{2}
\notag \\
&&+\left\vert a_{00}\right\vert ^{2}\left\vert b_{12}\right\vert
^{2}+\left\vert a_{01}\right\vert ^{2}\left\vert b_{12}\right\vert
^{2}+\left\vert a_{20}\right\vert ^{2}\left\vert b_{12}\right\vert
^{2}+\left\vert a_{21}\right\vert ^{2}\left\vert b_{12}\right\vert ^{2}
\notag \\
&&+\left\vert a_{01}\right\vert ^{2}\left\vert b_{20}\right\vert
^{2}+\left\vert a_{02}\right\vert ^{2}\left\vert b_{20}\right\vert
^{2}+\left\vert a_{00}\right\vert ^{2}\left\vert b_{21}\right\vert
^{2}+\left\vert a_{02}\right\vert ^{2}\left\vert b_{21}\right\vert ^{2}
\notag \\
&&+\left\vert a_{10}\right\vert ^{2}\left\vert b_{21}\right\vert
^{2}+\left\vert a_{12}\right\vert ^{2}\left\vert b_{00}\right\vert
^{2}+\left\vert a_{12}\right\vert ^{2}\left\vert b_{01}\right\vert
^{2}+\left\vert a_{12}\right\vert ^{2}\left\vert b_{20}\right\vert ^{2}
\notag \\
&&+\left\vert a_{12}\right\vert ^{2}\left\vert b_{21}\right\vert
^{2}+\left\vert a_{00}\right\vert ^{2}\left\vert b_{22}\right\vert
^{2}+\left\vert a_{01}\right\vert ^{2}\left\vert b_{22}\right\vert
^{2}+\left\vert a_{10}\right\vert ^{2}\left\vert b_{22}\right\vert ^{2}
\notag \\
&&+\left\vert a_{11}\right\vert ^{2}\left\vert b_{00}\right\vert
^{2}+\left\vert a_{11}\right\vert ^{2}\left\vert b_{02}\right\vert
^{2}+\left\vert a_{11}\right\vert ^{2}\left\vert b_{20}\right\vert
^{2}+\left\vert a_{11}\right\vert ^{2}\left\vert b_{22}\right\vert ^{2}
\notag \\
D_{2} &=&\left\vert a_{11}\right\vert ^{2}\left\vert b_{01}\right\vert
^{2}+\left\vert a_{21}\right\vert ^{2}\left\vert b_{01}\right\vert
^{2}+\left\vert a_{12}\right\vert ^{2}\left\vert b_{02}\right\vert ^{2}]
\notag \\
&&+\left\vert a_{22}\right\vert ^{2}\left\vert b_{02}\right\vert
^{2}+\left\vert a_{00}\right\vert ^{2}\left\vert b_{10}\right\vert
^{2}+\left\vert a_{20}\right\vert ^{2}\left\vert b_{00}\right\vert
^{2}+\left\vert a_{20}\right\vert ^{2}\left\vert b_{10}\right\vert ^{2}
\notag \\
&&+\left\vert a_{01}\right\vert ^{2}\left\vert b_{11}\right\vert
^{2}+\left\vert a_{21}\right\vert ^{2}\left\vert b_{11}\right\vert
^{2}+\left\vert a_{02}\right\vert ^{2}\left\vert b_{12}\right\vert
^{2}+\left\vert a_{22}\right\vert ^{2}\left\vert b_{12}\right\vert ^{2}
\notag \\
&&+\left\vert a_{00}\right\vert ^{2}\left\vert b_{20}\right\vert
^{2}+\left\vert a_{10}\right\vert ^{2}\left\vert b_{00}\right\vert
^{2}+\left\vert a_{10}\right\vert ^{2}\left\vert b_{20}\right\vert
^{2}+\left\vert a_{01}\right\vert ^{2}\left\vert b_{21}\right\vert ^{2}
\notag \\
&&+\left\vert a_{11}\right\vert ^{2}\left\vert b_{21}\right\vert
^{2}+\left\vert a_{02}\right\vert ^{2}\left\vert b_{22}\right\vert
^{2}+\left\vert a_{12}\right\vert ^{2}\left\vert b_{22}\right\vert ^{2}
\end{eqnarray}%
The coefficient$\ E$ in equation (\ref{24}) is given below,\
\begin{eqnarray}
E &=&\left\vert a_{00}\right\vert ^{2}\left\vert b_{00}\right\vert
^{2}+\left\vert a_{01}\right\vert ^{2}\left\vert b_{01}\right\vert
^{2}+\left\vert a_{02}\right\vert ^{2}\left\vert b_{02}\right\vert
^{2}+\left\vert a_{10}\right\vert ^{2}\left\vert b_{10}\right\vert ^{2}
\notag \\
&&+\left\vert a_{11}\right\vert ^{2}\left\vert b_{11}\right\vert
^{2}+\left\vert a_{12}\right\vert ^{2}\left\vert b_{12}\right\vert
^{2}+\left\vert a_{20}\right\vert ^{2}\left\vert b_{20}\right\vert
^{2}+\left\vert a_{21}\right\vert ^{2}\left\vert b_{21}\right\vert
^{2}+\left\vert a_{22}\right\vert ^{2}\left\vert b_{22}\right\vert ^{2}
\end{eqnarray}%
The coefficients$\ F_{i}$ in equation (\ref{25}) are given below,\
\begin{eqnarray}
F_{1} &=&\left\vert a_{21}\right\vert ^{2}\left\vert b_{01}\right\vert
^{2}+\left\vert a_{12}\right\vert ^{2}\left\vert b_{02}\right\vert
^{2}+\left\vert a_{22}\right\vert ^{2}\left\vert b_{02}\right\vert ^{2}
\notag \\
&&+\left\vert a_{20}\right\vert ^{2}(\left\vert b_{00}\right\vert
^{2}+\left\vert b_{10}\right\vert ^{2})+\left\vert a_{01}\right\vert
^{2}\left\vert b_{11}\right\vert ^{2}+\left\vert a_{21}\right\vert
^{2}\left\vert b_{11}\right\vert ^{2}  \notag \\
&&+\left\vert a_{02}\right\vert ^{2}\left\vert b_{12}\right\vert
^{2}+\left\vert a_{22}\right\vert ^{2}\left\vert b_{12}\right\vert
^{2}+\left\vert a_{10}\right\vert ^{2}\left\vert b_{20}\right\vert
^{2}+\left\vert a_{01}\right\vert ^{2}\left\vert b_{21}\right\vert ^{2}
\notag \\
&&+\left\vert a_{11}\right\vert ^{2}(\left\vert b_{01}\right\vert
^{2}+\left\vert b_{21}\right\vert ^{2})+\left\vert a_{02}\right\vert
^{2}\left\vert b_{22}\right\vert ^{2}  \notag \\
&&+\left\vert a_{12}\right\vert ^{2}\left\vert b_{22}\right\vert
^{2}+\left\vert a_{10}\right\vert ^{2}\left\vert b_{00}\right\vert
^{2}+\left\vert a_{00}\right\vert ^{2}\left\vert b_{10}\right\vert ^{2}
\notag \\
&&+\left\vert a_{00}\right\vert ^{2}\left\vert b_{20}\right\vert ^{2}  \notag
\\
F_{2} &=&(a_{12}b_{02}a_{10}^{\ast }b_{00}^{\ast }+a_{22}b_{02}a_{20}^{\ast
}b_{00}^{\ast }+a_{10}b_{00}a_{11}^{\ast }b_{01}^{\ast }  \notag \\
&&+\ a_{20}b_{00}a_{21}^{\ast }b_{01}^{\ast }+a_{11}b_{01}a_{12}^{\ast
}b_{02}^{\ast }+a_{21}b_{01}a_{22}^{\ast }b_{02}^{\ast }  \notag \\
&&+a_{02}b_{12}a_{00}^{\ast }b_{10}^{\ast }+a_{22}b_{12}a_{20}^{\ast
}b_{10}^{\ast }+a_{00}b_{10}a_{01}^{\ast }b_{11}^{\ast }  \notag \\
&&+a_{20}b_{10}a_{21}^{\ast }b_{11}^{\ast }+a_{01}b_{11}a_{02}^{\ast
}b_{12}^{\ast }+a_{21}b_{11}a_{22}^{\ast }b_{12}^{\ast }  \notag \\
&&+a_{02}b_{22}a_{00}^{\ast }b_{20}^{\ast }+a_{12}b_{22}a_{10}^{\ast
}b_{20}^{\ast }+a_{00}b_{20}a_{01}^{\ast }b_{21}^{\ast }  \notag \\
&&+a_{10}b_{20}a_{11}^{\ast }b_{21}^{\ast }+a_{01}b_{21}a_{02}^{\ast
}b_{22}^{\ast }+a_{11}b_{21}a_{12}^{\ast }b_{22}^{\ast })  \notag \\
F_{3} &=&a_{21}a_{20}^{\ast }(b_{01}b_{00}^{\ast }+b_{11}b_{10}^{\ast
})+a_{22}a_{21}^{\ast }(b_{02}b_{01}^{\ast }+b_{12}b_{11}^{\ast })  \notag \\
&&+a_{20}a_{22}^{\ast }(b_{00}b_{02}^{\ast }+b_{10}b_{12}^{\ast
})+a_{11}a_{10}^{\ast }(b_{01}b_{00}^{\ast }+b_{21}b_{20}^{\ast })  \notag \\
&&+a_{01}a_{00}^{\ast }(b_{11}b_{10}^{\ast }+b_{21}b_{20}^{\ast
})+a_{12}a_{11}^{\ast }(b_{02}b_{01}^{\ast }+b_{22}b_{21}^{\ast })  \notag \\
&&+a_{02}a_{01}^{\ast }(b_{12}b_{11}^{\ast }+b_{22}b_{21}^{\ast
})+a_{10}a_{12}^{\ast }(b_{00}b_{02}^{\ast }+b_{20}b_{22}^{\ast })  \notag \\
&&+a_{00}a_{02}^{\ast }(b_{10}b_{12}^{\ast }+b_{20}b_{22}^{\ast })
\end{eqnarray}%
\begin{eqnarray}
F_{4} &=&a_{21}a_{20}^{\ast }(b_{01}b_{00}^{\ast }+b_{11}b_{10}^{\ast
})+a_{22}a_{21}^{\ast }(b_{02}b_{01}^{\ast }+b_{12}b_{11}^{\ast })  \notag \\
&&+a_{20}a_{22}^{\ast }(b_{00}b_{02}^{\ast }+b_{10}b_{12}^{\ast
})+a_{11}a_{10}^{\ast }(b_{01}b_{00}^{\ast }+b_{21}b_{20}^{\ast })  \notag \\
&&+a_{01}a_{00}^{\ast }(b_{11}b_{10}^{\ast }+b_{21}b_{20}^{\ast
})+a_{12}a_{11}^{\ast }(b_{02}b_{01}^{\ast }+b_{22}b_{21}^{\ast })  \notag \\
&&+a_{02}a_{01}^{\ast }(b_{12}b_{11}^{\ast }+b_{22}b_{21}^{\ast
})+a_{10}a_{12}^{\ast }(b_{00}b_{02}^{\ast }+b_{20}b_{22}^{\ast })  \notag \\
&&+a_{00}a_{02}(b_{10}b_{12}^{\ast }+b_{20}b_{22}^{\ast })  \notag \\
F_{5} &=&a_{22}a_{20}^{\ast }(b_{02}b_{00}^{\ast }+b_{12}b_{10}^{\ast
})+a_{20}a_{21}^{\ast }(b_{00}b_{01}^{\ast }+b_{10}b_{11}^{\ast })  \notag \\
&&+a_{21}a_{22}^{\ast }(b_{01}b_{02}^{\ast }+b_{11}b_{12}^{\ast
})+a_{12}a_{10}^{\ast }(b_{02}b_{00}^{\ast }+b_{22}b_{20}^{\ast })  \notag \\
&&+a_{02}a_{00}^{\ast }(b_{12}b_{10}^{\ast }+b_{22}b_{20}^{\ast
})+a_{10}a_{11}^{\ast }(b_{00}b_{01}^{\ast }+b_{20}b_{21}^{\ast })  \notag \\
&&+a_{00}a_{01}^{\ast }(b_{10}b_{11}^{\ast }+b_{20}b_{21}^{\ast
})+a_{11}a_{12}^{\ast }(b_{01}b_{02}^{\ast }+b_{21}b_{22}^{\ast })  \notag \\
&&+a_{01}a_{02}^{\ast }(b_{11}b_{12}^{\ast }+b_{21}b_{22}^{\ast })
\end{eqnarray}

\begin{figure}[tbp]
\begin{center}
\vspace{-2cm} \includegraphics[scale=0.6]{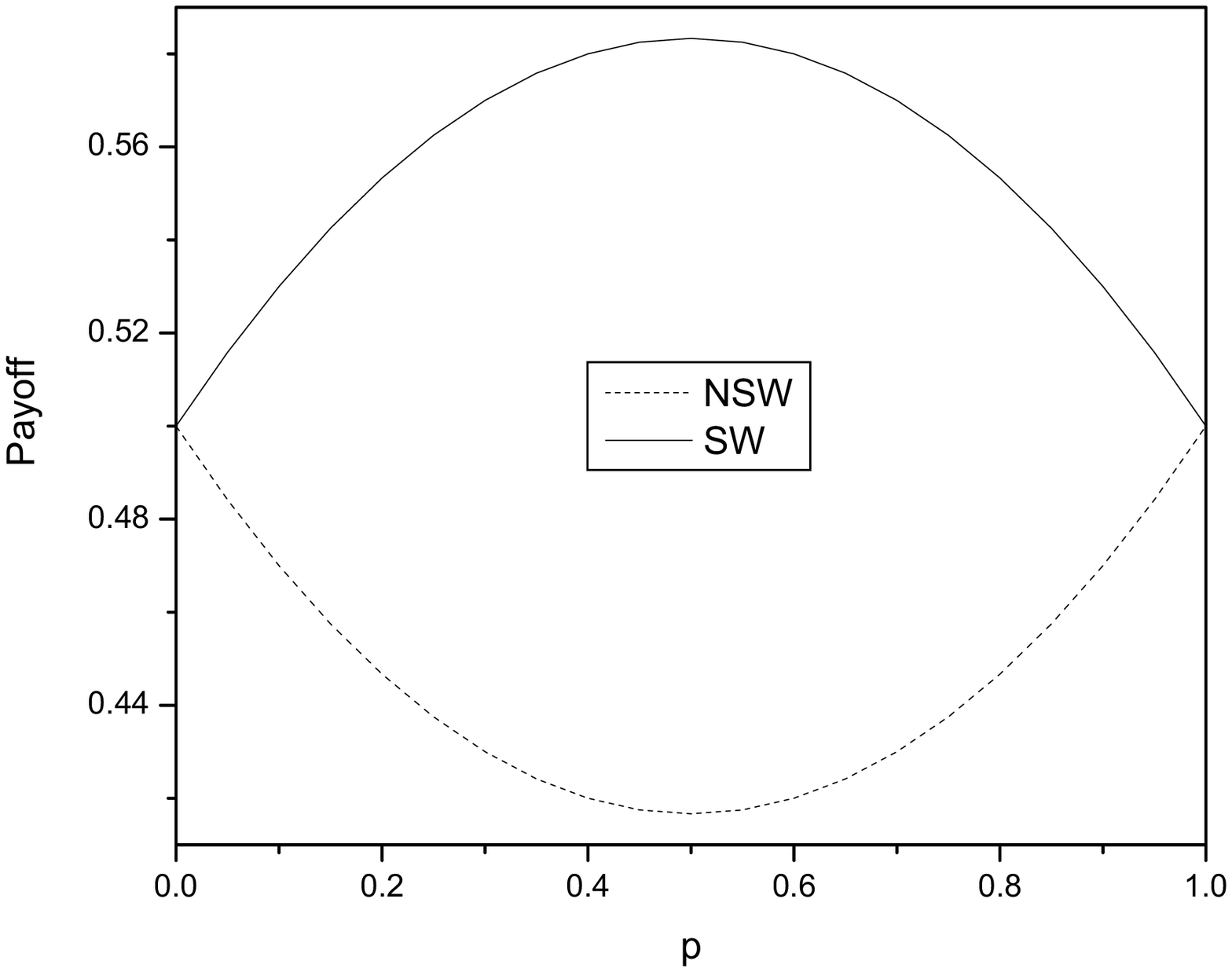} \\[0pt]
\end{center}
\caption{Bob's payoff is plotted as a function of decoherence parameter $p$
when Alice plays $H$ and Bob plays $I$ for amplitude damping channel$.$ The
solid line represents the Bob's payoff for $\protect\gamma =0$ and the
dashed line represents the Bob's payoff for $\protect\gamma =\frac{\protect%
\pi }{2}.$}
\end{figure}

\begin{figure}[tbp]
\begin{center}
\vspace{-2cm} \includegraphics[scale=0.6]{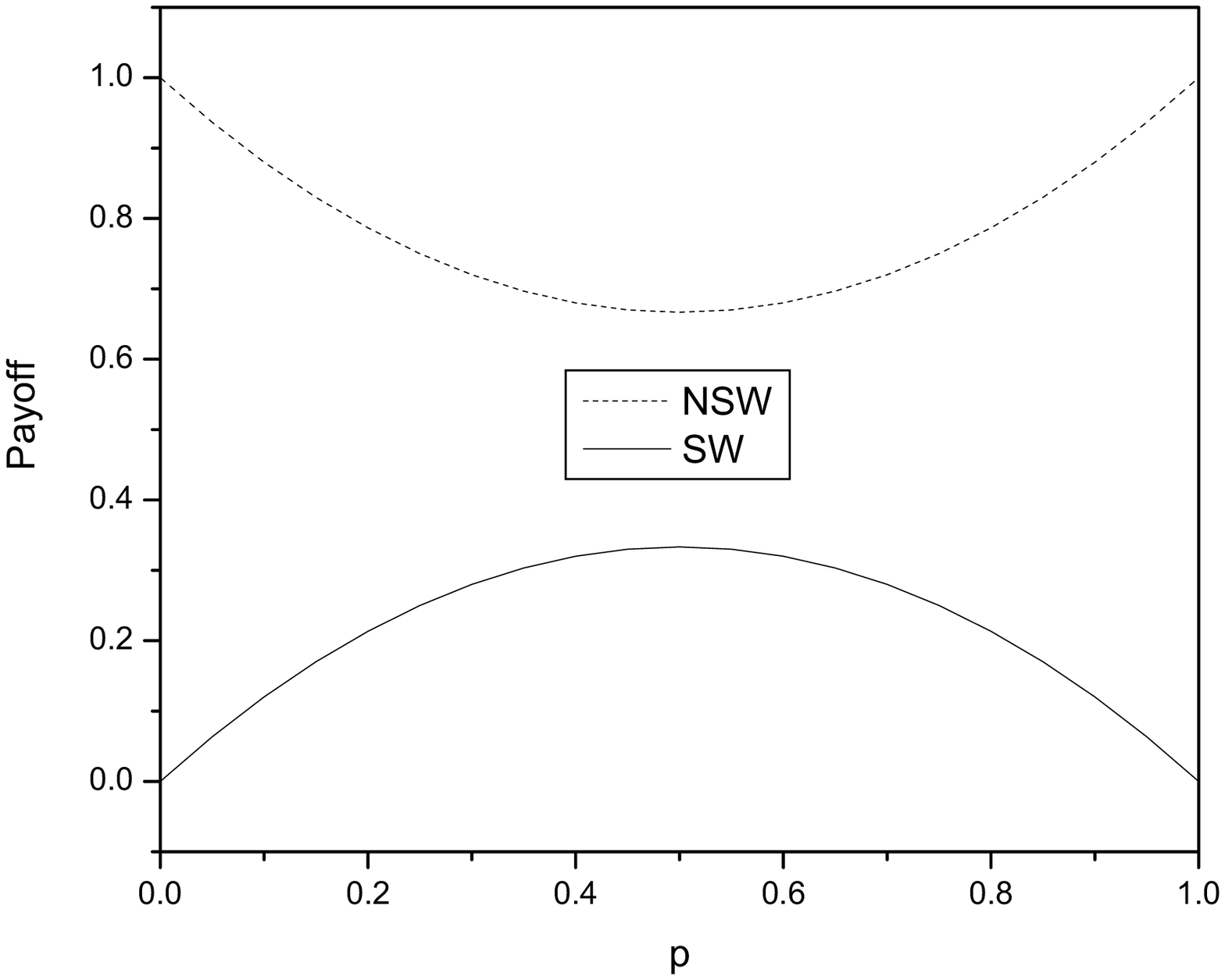} \\[0pt]
\end{center}
\caption{Bob's payoff is plotted as a function of decoherence parameter $p,$
when both Alice and Bob play $I$ , for amplitude damping channel$.$ The
solid line represents the Bob's payoff for $\protect\gamma =0$ and the
dashed line represents the Bob's payoff for $\protect\gamma =\frac{\protect%
\pi }{2}.$}
\end{figure}

\begin{figure}[tbp]
\begin{center}
\vspace{-2cm} \includegraphics[scale=0.6]{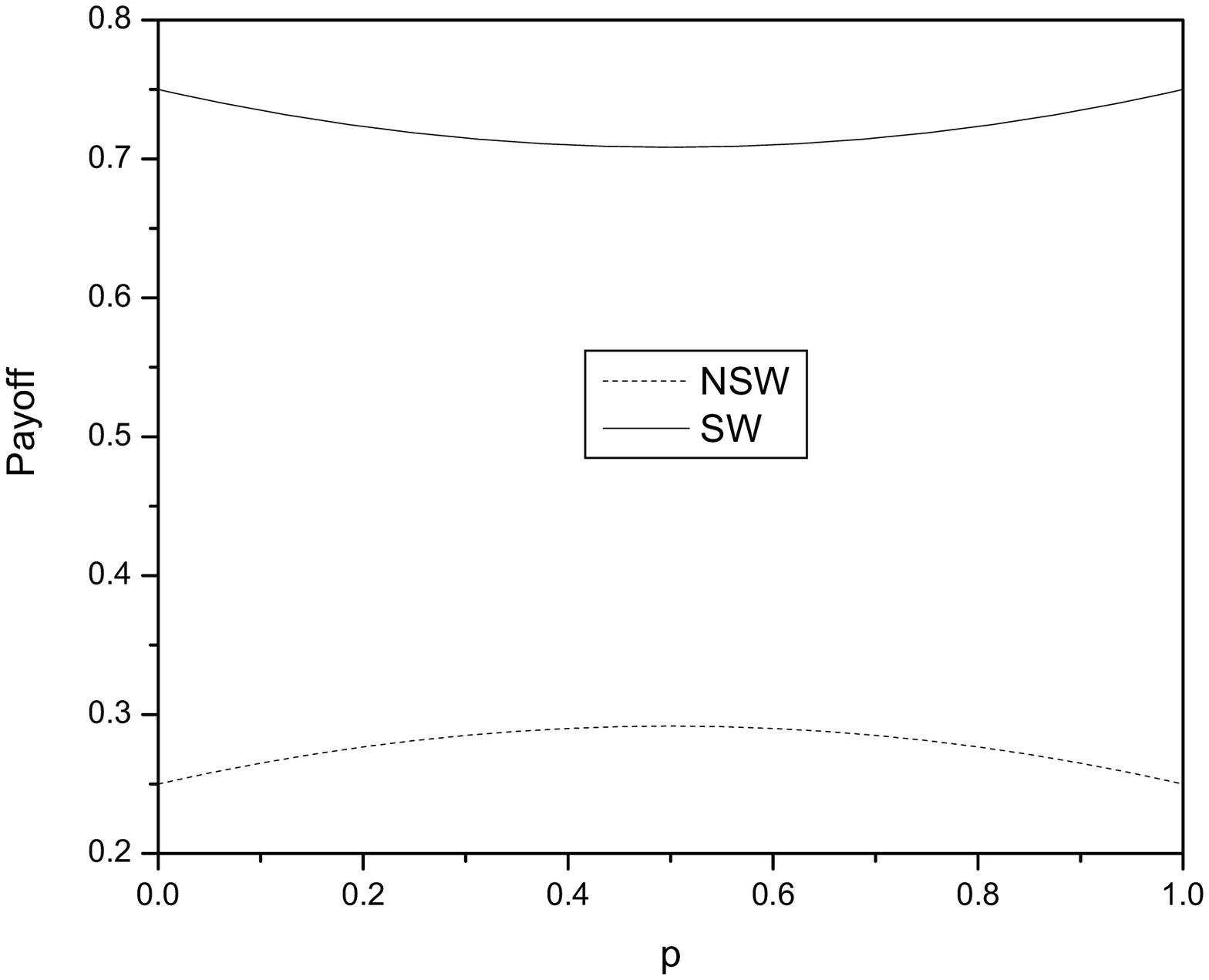} \\[0pt]
\end{center}
\caption{Bob's payoff is plotted as a function of decoherence parameter $p$
when Alice plays $H$ and Bob plays $M_{1}$ or $M_{2}$ for amplitude damping
channel$.$ The solid line represents the Bob's payoff for $\protect\gamma =0$
and the dashed line represents the Bob's payoff for $\protect\gamma =\frac{%
\protect\pi }{2}.$}
\end{figure}

\begin{figure}[tbp]
\begin{center}
\vspace{-2cm} \includegraphics[scale=0.6]{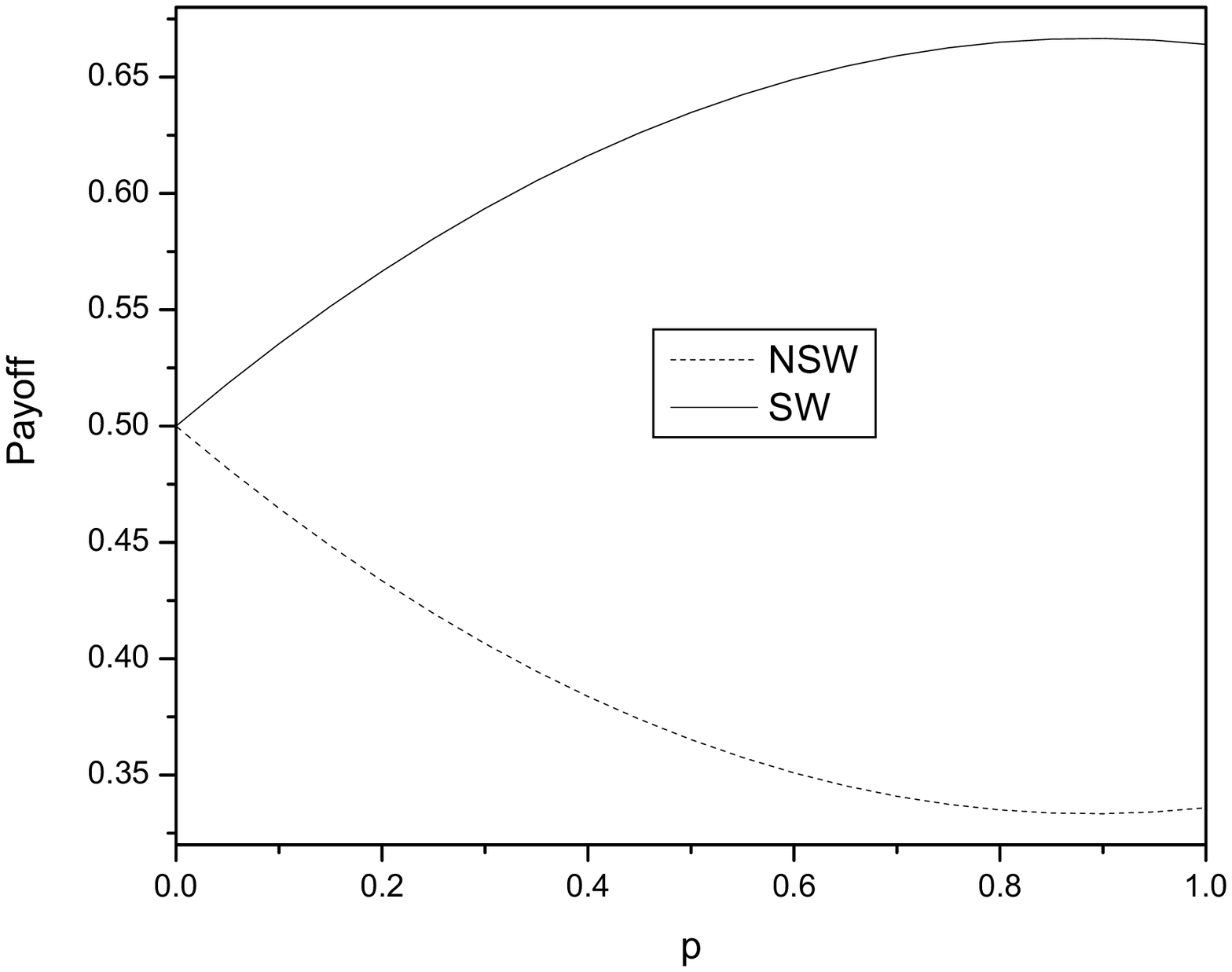} \\[0pt]
\end{center}
\caption{Bob's payoff is plotted as a function of decoherence parameter $p$
when Alice plays $H$ and Bob plays $I$ for depolarizing channel$.$ The solid
line represents the Bob's payoff for $\protect\gamma =0$ and the dashed line
represents the Bob's payoff for $\protect\gamma =\frac{\protect\pi }{2}.$}
\end{figure}

\begin{figure}[tbp]
\begin{center}
\vspace{-2cm} \includegraphics[scale=0.6]{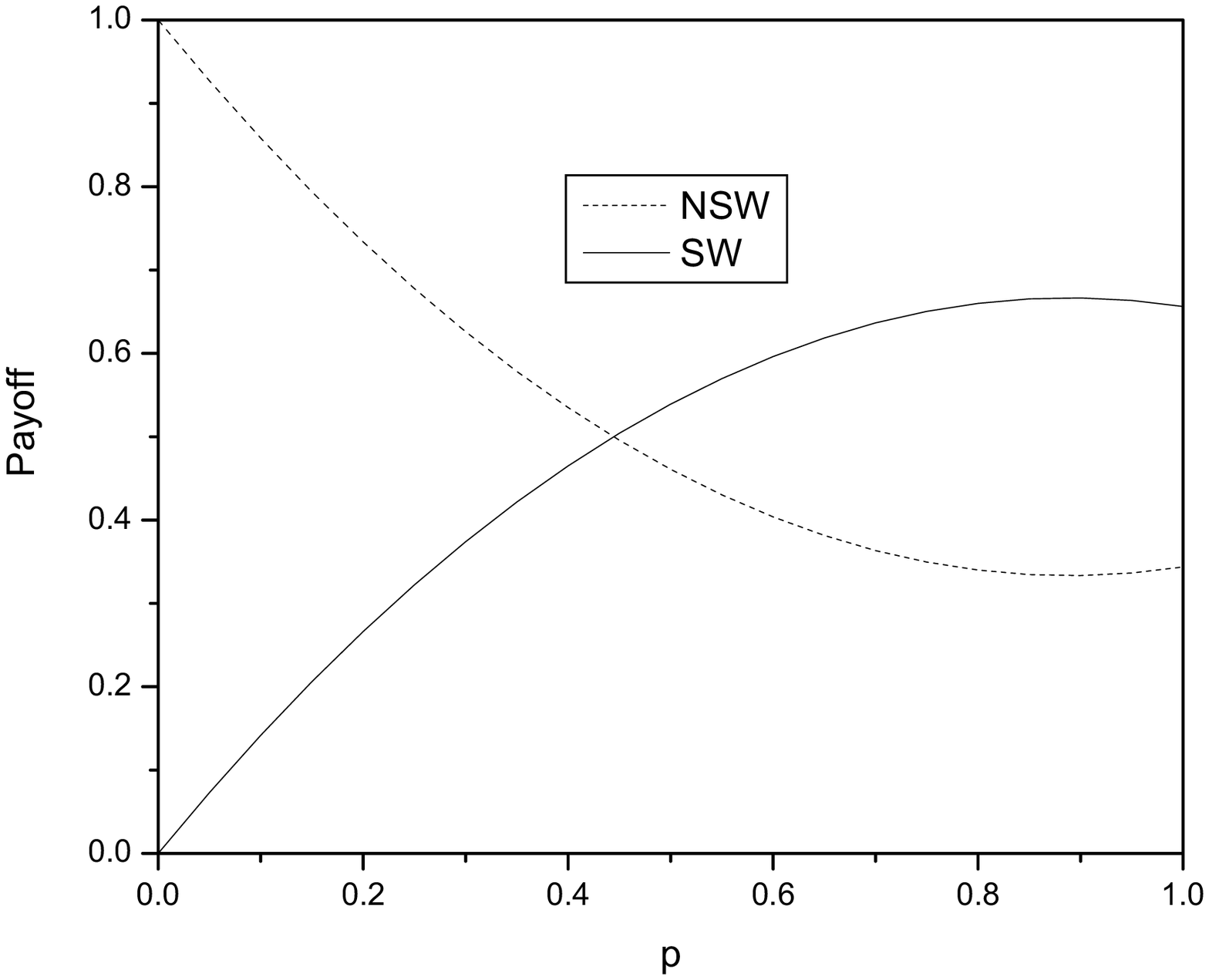} \\[0pt]
\end{center}
\caption{Bob's payoff is plotted as a function of decoherence parameter $p$
when both Alice and Bob play $I$ for depolarizing channel$.$ The solid line
represents the Bob's payoff for $\protect\gamma =0$ and the dashed line
represents the Bob's payoff for $\protect\gamma =\frac{\protect\pi }{2}.$}
\end{figure}

\begin{figure}[tbp]
\begin{center}
\vspace{-2cm} \includegraphics[scale=0.6]{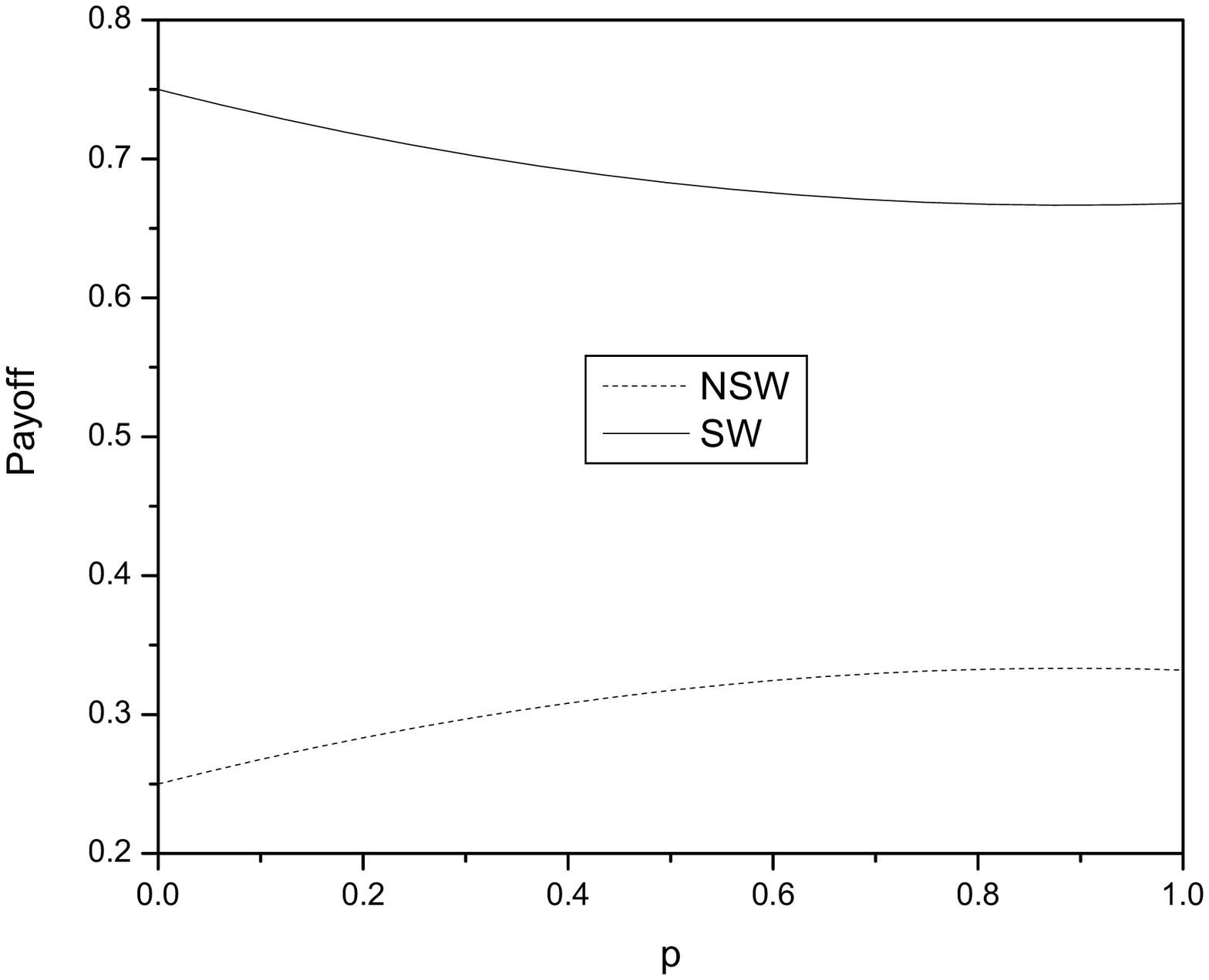} \\[0pt]
\end{center}
\caption{Bob's payoff is plotted as a function of decoherence parameter $p$
when Alice plays $H$ and Bob plays $M_{1}$ or $M_{2}$ for depolarizing
channel$.$ The solid line represents the Bob's payoff for $\protect\gamma =0$
and the dashed line represents the Bob's payoff for $\protect\gamma =\frac{%
\protect\pi }{2}.$}
\end{figure}

\end{document}